\documentclass[useAMS,usenatbib,usegraphicx]{mn2e}

\title[Differential Microlensing Measurements of Quasar Broad Line Kinematics]
{Differential Microlensing Measurements of Quasar Broad Line Kinematics in Q2237+0305}
\author[M. O'Dowd, N. F. Bate, R. L. Webster, R. Wayth and K. Labrie]{M. O'Dowd$^{1}$\thanks{E-mail:
matt@astro.columbia.edu}, N. F. Bate$^{2}$, R. L. Webster$^{2}$, R. Wayth$^{3}$, K. Labrie$^4$\\
$^{1}$Columbia University, New York, NY, USA\\
$^{2}$The University of Melbourne\\
$^{2}$Harvard\\
$^{3}$Gemini}
\begin{document}

\date{Accepted 2010 November 28}

\pagerange{\pageref{firstpage}--\pageref{lastpage}} \pubyear{2010}

\maketitle

\label{firstpage}

\begin{abstract}
The detailed workings of the central engines of powerful quasars remain a
mystery. This is primarily due to the fact that, at their cosmological
distances, the inner regions of these quasars are spatially
unresolvable. Reverberation mapping is now beginning to unlock the 
physics of the Broad Emission Line Region (BELR) in nearby,
low-luminosity quasars, however it is still unknown whether this gas
is dominated by virial motion, by outflows, or infall. The challenge
is greater for more distant, powerful
sources due to the very long response time of the BELR to
changes in the continuum. We present a new technique for probing
the kinematic properties of the BELR and accretion disk of high-z
quasars using differential microlensing, and show how substantial
information can be gained through a single observation of a
strongly-lensed quasar using integral field spectroscopy. We apply
this technique to GMOS IFU observations of the multiply-imaged quasar
Q2237+0305, and find that the observed microlensing signature in the
CIII] broad emission line favours gravitationally-dominated dynamics
over an accelerating outflow.

\end{abstract}


\section{Introduction}

Differential microlensing offers the promise of indirectly measuring
the spatial size and kinematics of different emitting regions in
quasars.   In this paper, a new observation of differential
microlensing between the continuum and different velocity components
of the CIII] broad line emission line in Q2237+0305 is presented.
These data are then used to constrain the kinematics of the broad line
gas. 

The core regions of quasars are unresolved at any wavelength, and so
physical understanding must be based on theoretical models
constrained by their observed spectral properties. The gas responsible
for the broad emission lines presents a particular challenge as there 
are relatively few measurements capable of even beginning to probe
its kinematical structure. As such, there is still no agreement
on even the simplest models of this structure.
Beyond the obvious width of the emission lines, which
indicate large Doppler velocities of tens of thousands of km/sec,
most of our current understanding has come from reverberation mapping
and from polarisation observations of emission lines.
The most recent
reverberation signatures have been interpreted as arising from
outflowing,  inflowing and virialised motions \citep{Denney09,
Bentz09}, while the polarisation studies have indicated an 
outflowing helical motion in the emission line gas
\citep{Young07}. Theoretical models have favoured outflowing winds (eg
\citealt{MurrayChiang97}), but alternative points of view are still under
active consideration (eg. \citealt{Gaskell09}).  

Differential microlensing provides an independent probe of the inner
structure of quasars. In quasars that are subject to strong
gravitational lensing, the BELR and potentially
the accretion disk are resolved on the spatial scale of the fine
magnification structure of the lens. Given a model for the
lensing galaxy, observation of differential magnification between
these components allows constraints to be placed on their sizes.

The use of this technique was first discussed by \citet{nem88}, who
specifically looked at the effect of a single low mass star on a range
of kinematical models for the broad emission line   region (BELR). A
key result of  this paper was that different parts of the emission
lines were differentially magnified, depending  on the kinematical
model: in general the smaller spatial regions showed the greatest
variations,  as would be expected. \citet{SW90} further considered the
problem, noting that, in the case of a  macro-imaged quasar, the
differences between the lines in different images can be used to test
whether particular emitting regions are being significantly
microlensed.  In particular, these  authors noted that Keplerian
motions in the BELR would be much easier to detect than infall  (or
presumably outflow).  A key improvement in the modelling compared to
\citet{nem88} was the use of microlensing  magnification  patterns to
model the possible statistical  variations for each macro-image.
\citet{SW90} also noted that differential microlensing would also
effect the redshift measured from a particular line.  More recently
\citet{Abajas07} and \citet{LewisAbata04} both considered
different signatures which might be induced in the observed structure
of the broad emission lines. 

There are at least four differential microlensing experiments which
could allow the measurement of the  physical parameters of the quasar
emission regions. Target-of-opportunity observations of a quasar
crossing a caustic  provide the cleanest imaging experiment
\citep{wwtm}.  The physical interpretation of a caustic crossing
event is straightforward.   However in order to trigger the
target-of-opportunity, regular  monitoring is required; and for a
reasonable annual probability of observing a caustic crossing,  more
than ten objects would need to be monitored.  Recently, monitoring
data has been used to fit  the size of the region emitting the quasar
continuum using Bayesian Monte Carlo methods \citep{koch04, Eigenbrod08b}.  These
analyses rely on detailed modeling  where a standard accretion disk
is convolved with microlensing networks.    A third and more
specialized possibility arises when two macro-images straddle a
caustic.  In a  surprising number of cases, the fluxes of the two
images differ, while theory predicts them to be  the same. Several
variables in the modeling can affect the relative fluxes, but the size
of the  emission region is the dominant factor \citep{cko, bww}.   
\citet{bww} have shown that measurements  of the anomalous fluxes can
be used to set limits on the size of the emission regions. 

The final  method uses spectroscopic data to compare
the shapes of emission lines and continuum spectra.  If  the broad
emission line region has ordered kinematical motions, then
differential magnification  may change the shape of the emission lines
of one image.  In the case of a macro-imaged quasar,  differences
between the line spectra of the images can then provide a diagnostic for
the kinematical motions of the broad emission lines. Such differences
have already been observed in several sources (e.g. \citealt{Keeton, Eigenbrod08b,
  Sluse, Hutsemekers}). In this work we
apply this last method to new integral field spectroscopy (IFS) of the 
multipy imaged quasar Q2237+0305 \citep{Huchra}.

Q2237+0305  is an ideal candidate for
differential microlensing experiments. The close proximity of the
lensing galaxy, at $z_d=0.0394$ compared to the quasar redshift of
$z_s=1.695$, yields a large projected Einstein Radius (ER) of   $\sim
2\times 10^{17}h_{70}^{1/2} (M/M_\odot)^{1/2}$cm in the source
plane. This ER, which characterizes the size-scale of magnification
fluctuations, is significantly larger than its estimated continuum
region size of $3\times 10^{16}h_{70}^{1/2} (M/M_\odot)^{1/2}$cm
\citep{Witt}, and it is similar to the upper size limit of its CIII]
and MgII BELR, determined from our previous observations of
differential microlensing in this source (\citealt{Wayth}, hereafter W05). This
indicates a high probability of differential microlensing between
continuum and broad lines, and also a reasonable probability of
differential microlensing {\it within} the BELR. 

Differential microlensing between the continuum and BELR in Q2237+0305 
has been observed many times. Microlensing within the broad line
itself has also been observed. Most notably, \citet{ecsma} have
conducted VLT monitoring of Q2237+0305 that spans
several years, and present high quality slit spectroscopy of both
continuum and BELR change, comparable to the data presented in this
paper. However IFS observations can provide more reliable spectra as
they allow more careful deblending of the small-separation lensed
images and lensing galaxy.

Interpretation of the differences between the spectra of the four
images requires a model for both the accretion disk and the spatial
and kinematical structure of the BELR.  Broadly, larger flux changes
are expected for regions closer to the accretion disk. The velocity
structure, projected along the line-of-sight to the quasar should
discriminate between models where the higher velocities are far from
the disk (due to an accelerating wind) and those where a high velocity
might be found close to the disk (due to Keplerian motion around the
central black hole). It is still not known which (if either) of these
two models best describes the kinematics of the BELR. 

The current work does not attempt to construct a fully consistent
model of the BELR, nor include a full photoionisation calculation.
Rather, a more limited question is addressed:  near the accretion disk
where the continuum is emitted, is the BELR gas moving relatively
quickly or slowly? Answering this question will help resolve the
dominant physical process in the inner BELR. 

In Section 2 of this paper we present the observational data, and the methods of
data reduction and spectra extraction. Section 3 describes the  flux
ratios for different emission lines and at different wavelengths for
the continuum. In Section 4 we present simple models to  describe the
BELR kinematics.  Section 5 describes the computation of the
microlensing simulations which are convolved with the  models.  In
Section 6, the results of comparing the model predictions with the
data are presented.  These are discussed in  Section 7, with the final
conclusions presented in Section 8. 

\section{Observational data}

Q2237$+$0305 was observed on the 27th of June 2006 with the GMOS
Integral Field Unit (IFU)
\citep{AllingtonSmith, Hook} on the Gemini South telescope (program ID:
GS-2006A-Q-14). 5$\times$15 minute exposures were taken in one-slit
mode using the B600 grating and with central wavelength 515 nm, 
giving a spectral range of 3975\AA\ to 6575\AA\ and a resolution of $R=1688$.
At Q2237$+$0305's redshift of 1.69, this covers the CIV and CIII]
broad emission lines.
Seeing conditions were excellent, with a PSF FWHM of 0\farcs6.

The GMOS IFU in one-slit mode gives a 5\arcsec$\times$3\farcs5 field
of view, with an additional 5\arcsec$\times$1\farcs25 field offset by 
$\sim$1~arcmin to measure the sky. Importantly, the field is 'filled',
with the science field consisting of 500 0\farcs2 
diameter hexagonal lenslets with abutting edges. The filled field
means that no interpolation between lenslets is required, which 
makes the determination of brightness ratios significantly more
accurate than in the case of non-filled IFUs.

\begin{figure}
 \includegraphics*[width=85mm]{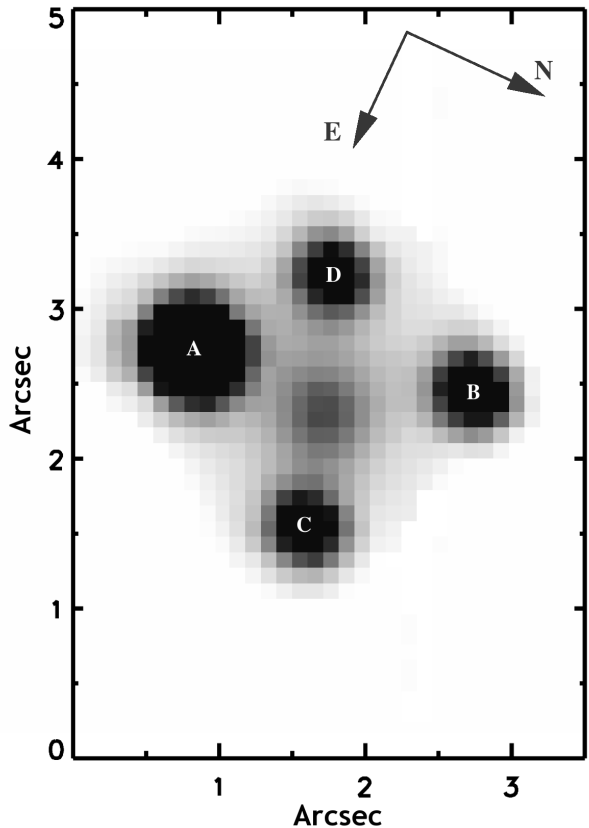}
  \caption{2-D image created from collapsing ADR-corrected IFU data
    cube. Letters indicate the standard designation of quasar images in
    the Einstein Cross.}
  \label{2dimage}
\end{figure}

\subsection{Data Reduction}

The data was reduced using the Gemini \textsc{IRAF} package v1.8,
under \textsc{IRAF} v2.12.
The standard GMOS IFU reduction recipe was followed for overscan
subtraction and trimming, bias subtraction, dark subtraction,
flatfielding, arc calibration, spectrum extraction, sky 
subtraction, and flux calibration. We used calibration files taken as
part of the standard Gemini Facility Calibration Unit (GCAL) suite.

Cosmic ray removal was performed
separately on the reduced images before spectrum extraction, using
\textsc{CRREJ} algorithm via \textsc{IMCOMBINE}.  \textsc{CRREJ}
performs iterative rejection of unusually high pixels, accounting for
the noise properties of the image.
In W05 we found that this method provided far superior cosmic ray
rejection to that provided by \textsc{GSCRREJ}, and demonstrated the advantage 
of taking multiple GMOS IFU exposures at the same offset position, as
opposed to using an extensive dither pattern.
We converted the hexagonal image arrays into rectangular arrays with
0\farcs1 square pixels using \textsc{GFCUBE}. 

Atmospheric differential refraction (ADR) was found to cause a 0\farcs8
shift over the wavelength range. This was corrected by first computing
the shifts relative to the image position at the central wavelength of
CIII] using cross-correlation, then fitting a physical model of ADR
\citep{Schubert} to these shifts. The cube was corrected according to the
best-fit ADR model.

Figure \ref{2dimage} shows the ADR-corrected data cube collapsed into
a 2-D image. The four quasar images and the core of the lensing galaxy
are well-resolved. 

\subsection{PSF Construction and Extraction of Image Spectra}
\label{psffits}

\begin{figure*}
  \includegraphics[width=100mm]{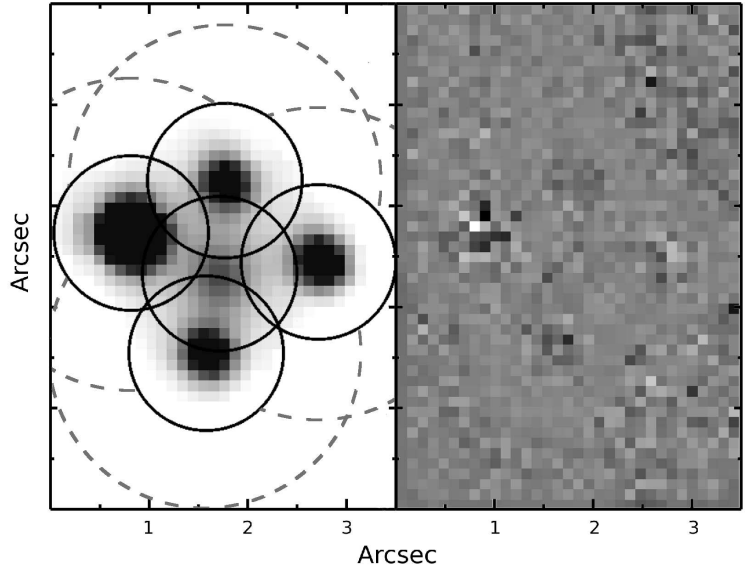}
  \caption{{\it Left}: Each spectrum is extracted from
    a 0\farcs8 aperture centred on the relevant quasar image or the
    lensing galaxy core (solid circles), but excluding regions overlapping any other
    0\farcs8 aperture. The model PSF is constructed with contributions from
    each quasar image, using 1\farcs6 apertures
    (dashed circles), and again excluding regions within 0\farcs8 of
    other images or the galaxy core (see Sect.~\ref{psffits} for details).
   {\it Right}: the residual map around the CIII] line after the
    best-fit subtraction of the constructed PSF from each quasar image
    and galaxy core. The median of the absolute noise-weighted
    residuals is $\sim 0.2\sigma$, with a range of $\pm \sim2\sigma$}.
  \label{psfext}
\end{figure*}

\begin{figure*}
  \includegraphics[width=160mm]{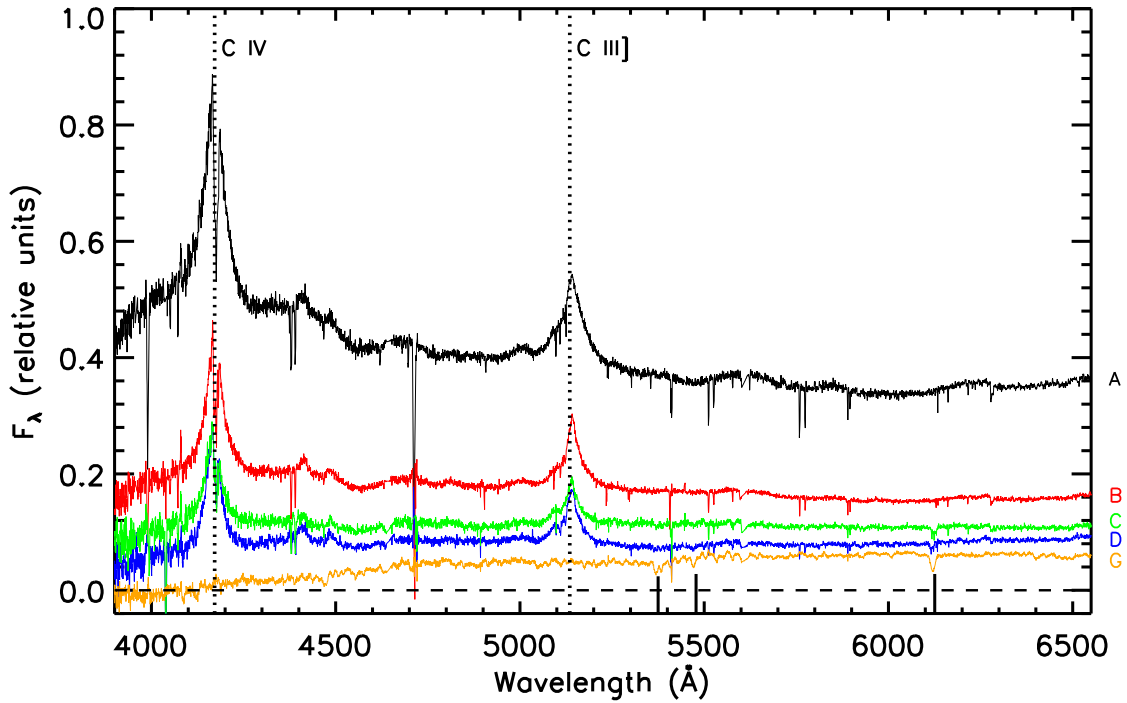}
  \caption{Extracted spectra for the four quasar images and
    the lensing galaxy in Q2237+0305:
  black: image A, red: image B, green: image C, blue: image D, orange:
  lensing galaxy. Dashed
  lines mark positions of CIV and CIII] broad emission lines. Solid
  vertical lines indicate positions of galaxy absorptions lines, from
  left to right: Mgb, FeII, and NaD}
  \label{spec}
\end{figure*}

The excellent seeing allows us to extract the spectum of each quasar
image with high precision. Nonetheless, there is some blending between
adjacent quasar images and with the core of the lensing galaxy, and so
deblending is required. 

To deblend, we perform least-squares fitting of the spatial frame
at each wavelength element. With positions determined from archival HST
imaging, we fit a reconstructed Point Spread Function (PSF; see below)
to all quasar images and the galaxy core using $\chi^2$ minimization. 
For each quasar image, we perform PSF-subtraction of all other
quasar images and the galaxy core using the best-fit amplitudes. 
We then determine the flux for each image, at each wavelength element,
within an integration aperture, normalized by the fraction of the
model PSF within this aperture. This aperture is a 0\farcs8 radius
circle centred on the relevant quasar image or galaxy, excluding regions within
0\farcs8 of any other quasar image or the galaxy core.

The PSFs used in these fits are determined using an iterative
process. As the PSF is wavelength-dependent, we calculate ten different
PSFs across the wavelength range, each from a $\sim$100\AA\ wide stacked
frame. The PSF extraction proceeds as follows:

{\bf \it a.} Take the azimuthal average for each quasar image within
1\farcs6 circles, excluding regions within 0\farcs8 
of any other quasar image or of the galaxy core. This yields four 1-D PSF
profiles. We splice these together to construct a single PSF profile
and convert this into a azimuthally symmetric 2-D PSF.

{\bf \it b.} Perform least-squares fitting to
obtain a first estimate of quasar image and galaxy core amplitudes.

{\bf \it c.} Create four frames; in each one, three of the
quasar images and the galaxy core are subtracted according to the
best-fit normalizations from step {\it b}, leaving a single quasar image
remaining. 

{\bf \it d.} Use these four frames to splice together a 2-D estimate of
the PSF using the same regions defined in step {\it a}.

{\bf \it e.} Repeat steps {\it c} through {\it d} using the 2-D PSF.

{\bf \it f.} Repeat until the PSF converges, with RMS differences between iterations of $<0.1$\%.

Using this process, the PSF converged within 5 iterations for all ten 100\AA\
wavelength bins, and was stable for further iterations.

Figure~\ref{psfext} illustrates the regions used for construction of the PSF and spectral extraction
(left), and the residuals after final subtraction for the spectral region
of the CIII] line. Figure~\ref{spec} shows the extracted
spectra of the four quasar images. 

The use of the PSF to model the core of the lensing galaxy introduces
some error.  If we assume a de Vaucouleurs' effective radius of
3\farcs1 \citep{swl98}, we find that the difference in
galaxy flux between the de Vaucouleur and PSF models in the
integration aperture of images A and B is less than 1\% of the quasar
image flux in this region. As the galaxy spectrum is reasonably smooth
in the region of the CIII] line, any residual galaxy flux is largely removed in the
continuum subtraction (see Section~\ref{sr}).

\section{Flux and Spectral Ratios}

In a multiply-imaged quasar, differential microlensing is
characterized by a variation in the magnification levels of different
spectral components in a single lens image, compared to the ``true'',
unmicrolensed quasar spectrum. This effect is most easily observed in 
the ratio of lens images. In the absense of microlensing,
we expect emission line and continuum flux ratios to be equivalent to
each other and to agree with the macro-model flux ratios, and we also
expect flat, featureless spectral ratios.
Gravitational lensing of a point source is achromatic, and so the
spectral differences resulting from differential microlensing can be identified
independently of the lens model. However it is important to consider
the other processes that can also lead to spectral differences between
lens images.

\noindent
\subsection{Differential Extinction and Intrinsic Variability}

It is possible that extinction properties differ significantly between
the lines of sight of each quasar image; both within the lensing
galaxy and in intervening absorbing systems. However 
in this analysis we are concerned with the relative strengths of the
continuum and BELR within a small wavelength range (the width of the
emission line), and so differential extinction will not be a
significant effect. Nonetheless, we correct for extinction using the
$A_V$ values of \citet{Agol}, which uses the average Milky Way
extinction curve to obtain $A_V$ for images A of 0.88$\pm$0.21, B of
0.84$\pm$0.20, C of 1.30$\pm$0.31, and D of 1.15$\pm$0.27. 

Intrinsic variation within the quasar coupled with 
time delays between lens images can result in time-varying spectral
differences, possibly duplicating the effect of differential
microlensing if the continuum undergoes significant variation over a
timescale smaller than that of the time delay. In the case of
Q2237+0305 the time delay predicted by modeling is less than a day \citep{kf88},
and Chandra observations measure a delay between images A and B of
$\sim 3$ hours \citep{Dai}. This is sufficiently smaller than the
estimated accretion disk light-crossing time of $\sim 6$~days to
eliminate the possibility of significant spectral differences due to
intrinsic variation.

\begin{figure}
  \includegraphics*[width=85mm]{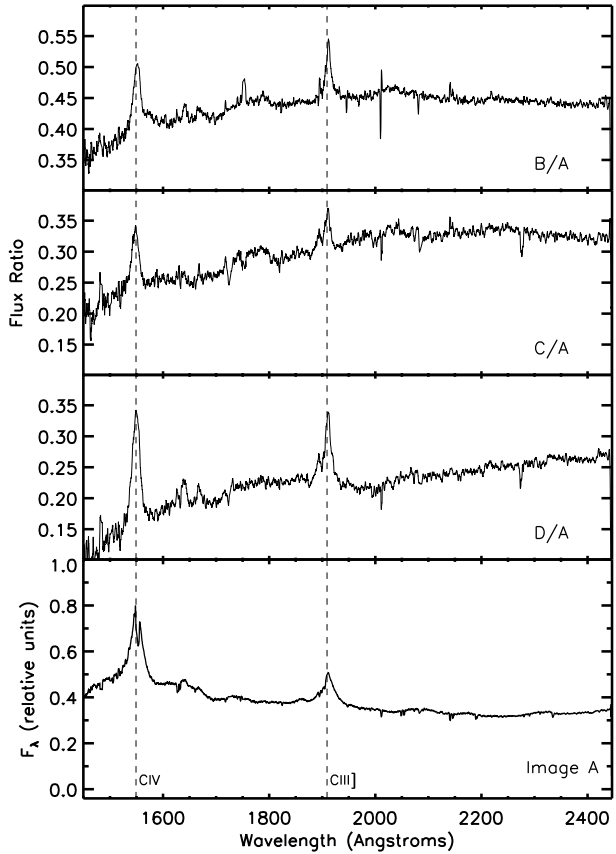}
  \caption{Ratios of spectra for images B, C, and D to image
    A. Residual features at locations of broad emission lines CIV and
    CIII] (dashed lines) reveal differential microlensing in which the
    continuum in image A is magnified relative to the BELR.}
  \label{specratios}
\end{figure}

\subsection{Flux Ratios}
\label{fr}

Table~\ref{fluxratios} gives the flux ratios in the CIV and CIII]
emission lines and in the continuum. Line ratios use the integrated,
continuum-subtracted line fluxes, while continuum ratios use the the continuum level at the
central wavelength of each line, determined by a linear interpolation
from either side of each line. From these ratios it is apparent that image A is
magnified relative to the other images. 

\begin{table}
\caption{Flux ratios for CIII] and CIV broad emission lines and for
  the continuum at the central wavelengths of these lines. Mid IR flux
  ratios from \citet{Agol}, radio flux ratios from \citet{Falco},
  H$\beta$ from \citet{Metcalf04}.}
\label{fluxratios}
\begin{tabular}{lccl}
\hline
Region        & B/A & C/A & D/A \\
\hline
CIII] line &  $0.569 \pm 0.022$ & $0.643 \pm 0.025$ & $0.630 \pm 0.026$ \\
CIII] cont. & $0.431 \pm 0.026$ & $0.443 \pm 0.027$ & $0.281 \pm 0.018$ \\
CIV line & $0.473 \pm 0.00;020$ & $0.454 \pm 0.020$ & $0.422 \pm 0.021$ \\
CIV cont. & $0.375 \pm 0.034$ & $0.347 \pm 0.033$ & $0.203 \pm 0.025$ \\
\hline
Mid IR     & $1.11 \pm 0.11  $ & $0.59 \pm 0.09 $  & $1.00 \pm 0.10 $ \\
8 GHz      & $1.08 \pm 0.27 $  & $0.55 \pm 0.21 $  & $0.77 \pm 0.23 $ \\
$H\beta$ line & $0.376 \pm 0.007$ & $0.387 \pm 0.007 $ & $0.461 \pm 0.004$\\
\hline
\end{tabular}
\end{table}

We also see differential microlensing between the BELR and the
continuum, with the continuum magnified to a greater extent than
the BELR. This is expected: given any level of differential
microlensing, the smaller emission region should exhibit the most
extreme deviation from the macro-model magnification. Using a method
similar to that used in W05, we find a size limits for the
CIII] BELR of $< 0.066h_{70}^{1/2}$~pc and for the continuum
emission region of $< 0.059h_{70}^{1/2}$~pc, both at the 95\%
confidence level. Note that, while the limit on the BELR size is
similar to that found from our 2002 data, the continuum limit is
less tight than the limit of $< 0.02h_{70}^{1/2}$~pc reported in
W05. This is because the earlier microlensing event resulted in stronger
magnification of the continuum --- possible only for smaller
accretion disk sizes. Our continuum size limit is also consistent
with, but less tight than, the limits found by \citet{PK} and
\citet{Eigenbrod08b}. 

We omit the CIV line from this analysis due to the difficulty in
performing an accurate continuum subtraction at the far blue end of the
spectrum.

\subsection{Spectral Ratios}
\label{sr}

Figure~\ref{specratios} shows the spectral ratios of each quasar
image relative to image A. The presence of features around 1549\AA\
and 1909\AA\ illustrates the differential microlensing between BELR and
continuum. The narrowness of these features suggest that there may be
differential microlensing across the velocity structure of the
line. In the case of a static magnification difference between line
and continuum we would expect a broad feature covering the
extent of the emission line. We also note that the continuum spectrum
of image A is bluer than that of the other images, indicating possible 
differential microlensing across the accretion disk.

\begin{figure*}
  \includegraphics[scale=0.6]{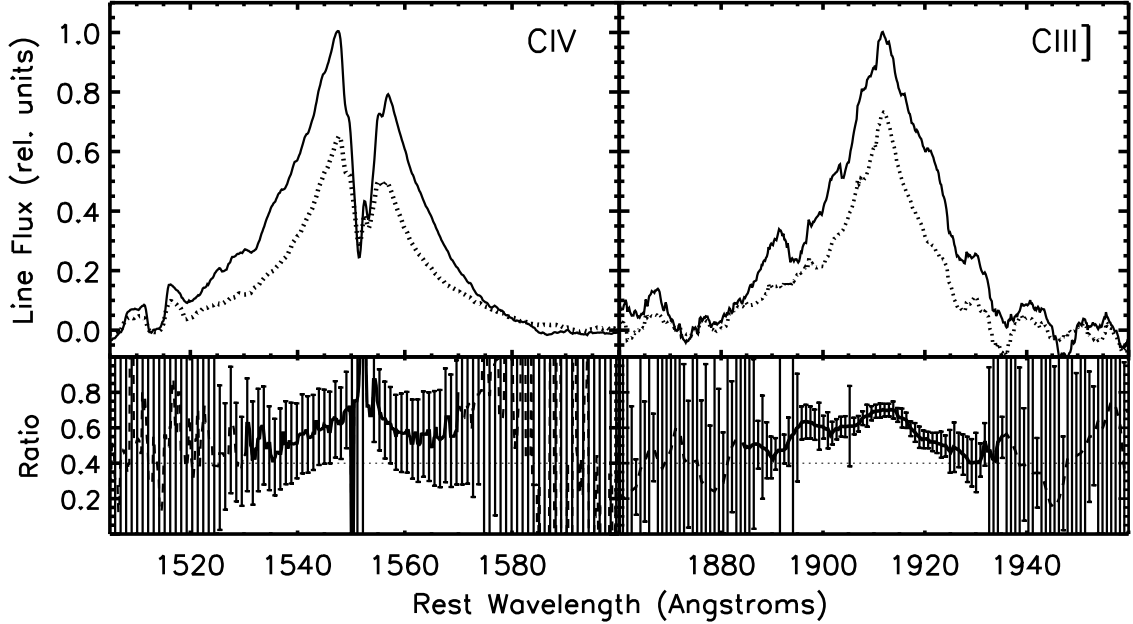}
  \caption{{\it Upper:} continuum-subtracted CIV and CIII] emission
    lines for images A ({\it solid}) and B ({\it dotted}). 
    {\it Lower:} B/A ratios using continuum-subracted emission
    lines. Solid lines show regions with significant line flux, while
    dashed lines show regions heavily dominated by noise. Dotted line
    shows the B/A continuum flux ratio.}
  \label{lineratios}
\end{figure*}

To investigate differential microlensing within the broad line region, we look at the
continuum-subtracted emission line ratios. The continuum spectrum at
each line is approximated as the linear fit to a 120\AA-wide region
centered on the line, masked in the central 80\AA-wide region. 
Figure~\ref{lineratios} shows the continuum-subtracted CIV and CIII]
lines for images A and B (upper plots) and the
continuum-subtracted line ratios (lower plots), which are equivalent
to emission line magnification ratios.

Errors in the spectral line ratios are derived from uncertainties in the
PSF fits (see Sect.~\ref{psffits}), which take into account errors in
subtraction of other images and of the lensing galaxy. We include the
error in continuum subtraction, determined conservatively from the
error in the linear fit around each line --- both from the fit itself
and from the variation in the continuum level from fitting to different
spectral regions. For the simple analysis to
follow, we don't directly calculate the error due to contamination by
any iron lines beneath the broad emission lines. Instead we assume that this
uncertainty is small compared to the uncertainty in the continuum. We
have no direct means to test this assumption.

Both the CIV and the CIII] line ratios reveal a symmetric trend in the
magnification ratio about the line centre.  
This trend is highly significant for the CIII] line. While the trend
is apparent for the CIV line, the difficulty in continuum subtraction
for this line (see Sect.~\ref{fr}) dramatically increases the
uncertainty in the spectral line ratios, and so we restrict further
analysis of this differential microlensing feature to the CIII] line.

At low velocities near the line centre, the CIII] broad line in image A is
magnified compared to the macro-model expectation: $B/A \sim 0.65$
compared to the expected $B/A \sim 1.0$. As line velocity increases
this magnification ratio drops further, towards the continuum ratio of 
$B/A \sim 0.4$. 
The observed trend suggests that the lower velocity broad line emission samples
a region of the caustic structure closer to the average macro-model magnification
level for image A than both the higher velocity gas and the
continuum. This in turn suggests that size of the emission region
approaches that of the continuum region with increasing gas
velocity. Note that this feature cannot be explained by differential
microlensing between the narrow line region versus the BLR as the
width of the feature exceeds 20\AA, or $\pm$1500~km/s.
In the next section we analyse
this microlensing signature in the context of simple models of BELR
kinematics.

\section{Models of the BELR}
\label{models}

We test simple models for two of the physical processes that are
considered to be the primary candidates as the dominant drivers of
BELR kinematics: an outflow driven by radiation pressure and orbital
motion driven by gravity. 
The aim of these models is to simulate the expected
wavelength-dependent surface brightness distributions for these
models, and then, by convolving these with simulated magnification
maps for Q2237+0305, to determine the expected magnification ratios
as a function of wavelength, which we will compare to the observed
microlensing signature.

\subsection{The Outflow Model}

For our simple outflow model we assume a radial wind in twin conical
shells. Drawing on the wind model of \cite{elvis},
we consider optically thick clouds driven outwards by radiation
pressure, with each cloud sufficiently small cross-sectional area to
allow us to ignore 
self-shielding. Under these assumptions we may use the derivation of
\cite{capriotti}, which gives the acceleration of a cloud due to
radiation pressure as:

\begin{equation}
a_{rad}=\frac{A_c L}{4\pi c r^2 M_c}
\end{equation}

\noindent
where $A_c$ is the cross-sectional area of the cloud, $M_c$ is its
mass, $L$ is the total ionising luminosity, and $r$ is the distance
from the ionising source. Velocity as a function of radius is therefore:

\begin{equation}
v dv=\frac{A_c L}{4\pi c r^2 M_c} dr
\end{equation}

\noindent
So for a starting velocity $v_0$ at an initial radius $r_0$ we have:

\begin{equation}
v(r)=\sqrt{(\frac{K+r_0 v_0^2}{r_0}-\frac{K}{r})}
\label{vel}
\end{equation}

\noindent
where $K=A_c L/2\pi c M_c$. And the terminal velocity, at
$r\rightarrow\infty$, is is $v_T=\sqrt{(K+r_0 v_0^2)/r_0}$. 

At a given projected distance from the centre, $x$, we know $v_{los}=v
\sqrt{1-\frac{x^2}{r^2}}$.
So from equation~\ref{vel}, the radius at which a given $v_{los}$
occurs along a given line of sight satisfies the cubic:



\begin{equation}
f(a)=a^3-\frac{a^2}{r_0}-\frac{a}{x^2}+(\frac{1}{x^2 r_0}-\frac{v_{los}^2}{x^2 K})=0
\label{cubic}
\end{equation}

\noindent
where $a=1/r$. This function yields a single physical solution for
the distance from the ionising source of clouds with projected velocity
$v_{los}$ at a given projected distance from the centre, $x$.

\noindent
We wish to determine the expected surface brightness distribution produced by
such a wind as a function of wavelength. 
If we assume that broad line clouds re-emit a constant fraction of the
incident ionising flux in the emission line, then the surface
brightness of an emission line at $\Delta \lambda$ from its rest
wavelength and at projected distance $x$ from the centre is: 

\begin{equation}
\mu_{x, \Delta \lambda}\propto \frac{d \delta x^2 \delta z}{r^2}
\end{equation}

\noindent
where $d$ is the number density of broad-line clouds at $r$, $\delta
x$ is the size of the spatial bin, and 
$\delta z$ is the distance along the line of sight over which
the emission line is shifted into our spectral resolution bin, 
assumed to be small compared to $r$.
For a spherical outflow, if we assume cloud conservation (i.e. their lifetime is long
compared to the flow timescale), then $d$ satisfies: $d \propto 1/(v r^2)$.
The size of projected spatial bins, $\delta x$, is constant, and so:

\begin{equation}
\mu_{x, \Delta \lambda}\propto \frac{\delta z}{v r^4}
\label{sb}
\end{equation}

So combined with equation~\ref{vel} we have surface brightness at
$x$ and $\Delta \lambda$ as a function of the distance from the
source at which that emission is produced. For the purpose of the
model it is useful to parameterize $K$ in terms of $r_0$, $v_0$, and $v_T$:

\begin{equation}
\mu_{x, \Delta \lambda}\propto \frac{\delta z}{\sqrt{(v_0^2-v_T^2)\frac{r_0}{r}+v_T^2}}
\end{equation}

\noindent
where $r$ and $\delta z$ are calculated using equation~\ref{cubic}. 

We include both axial and equatorial obscuration with 
varying opening angles to simulate constrained winds or obscuration by
jet and disk/torus. The final model of a three-dimensional
axially symmetric radial wind is parameterized by:
launching radius, terminal velocity (which
encompasses cloud properties and continuum luminosity), orientation of
the axis to the line of sight,
and the opening angle of axial and equatorial black-out regions.

\begin{figure}
  \includegraphics*[width=90mm]{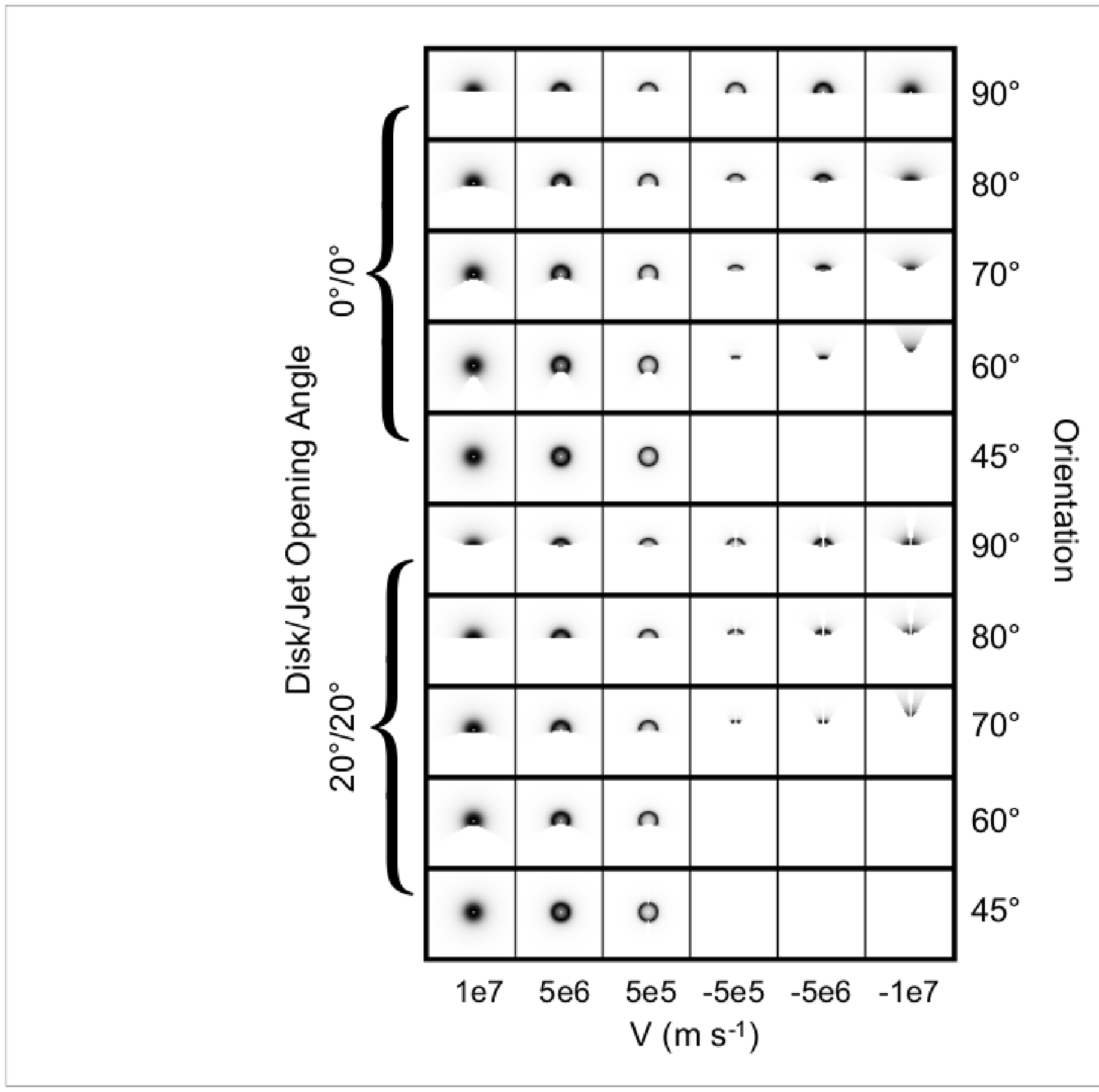}
  \caption{Outflow model surface brightness maps for a selection of
    the explored parameter space. The depicted models have launching
    radius $r_0=1\times10^{16}$~cm and a range of disk/jet opening
    angles and orientations of the axis to the line of sight.
    For each parameter
    set, a small selection of the calculated velocity slices are
    shown, with negative velocities indicating the receding
    outflow. The surface brightness peak of the receding outflow is
    increasingly displaced relative to the approaching outflow as
    velocity increases. An obscuring disk and jet blacks out a
    larger fraction of the receding wind as orientation
    angle decreases and velocity increases, completely blacking out even
    lower velocity gas at angles below 60\degr to 70\degr.
    Each frame size is $3\times10^{17}$~cm.}
  \label{windsb}
\end{figure}

Figure~\ref{windsb} shows an example set of surface brightness
distributions for the wind model. As can be seen, significant
asymmetries arise for axis orientations not perpendicular to the
line of sight. Any obscuring disk or jet quickly blacks out the entire 
receding wind as velocity 
increases, and this obscuration increases with smaller orientation
angles. Interestingly, the size scale of the surface brightness
distribution does not change dramatically with gas velocity, although
the red- and blue-shifted high velocity gas is offset in the plane of
the sky for low orientation angles. Such offsets are likely to lead to
strong differential microlensing, and hence asymetries between red-
and blue-shifted line ratios.

\subsection{The Orbital Model}

To study the expected microlensing signature of an orbital BELR, we
assume a purely Keplerian model consisting of randomly-oriented 
circular orbits. In this case the analog of equation~\ref{cubic}:

\begin{equation}
f(a)=x^2 a^3-a^2-\frac{v_{los}^2}{K'}=0
\label{kcubic}
\end{equation}

\noindent
where $a=\frac{1}{r^2}$, $K'=\frac{G M_{BH}}{2\pi^2}$, and as previously
$x$ is the projected distance from the centre and $v_{los}$ is the
line-of-sight velocity that we're interested in. The solution to this
cubic gives us the radius at which $v_{los}$ occurs for the current $x$.

We follow the same approach as for the wind model, except that the 
cloud number density can't be constrained by simple conservation 
arguments, and so we need to choose a density gradient, $d\propto
r^{\alpha}$. We choose a value of $\alpha = -1$, in the mid-range of
the mid-range of likely values (e.g.~\citealt{Goad}).

The surface brightness distribution of a purely Keplerian BELR is then:

\begin{equation}
\mu_{x, \Delta \lambda}\propto K' \delta z r^{\alpha-2}
\end{equation}

The final model of a spherical BELR comprised of random circular
Keplerian orbits is parameterized by:
black hole mass, density gradient, orientation of
the axis to the line of sight,
and the opening angle of axial and equatorial black-out regions.

\begin{figure}
  \includegraphics*[width=90mm]{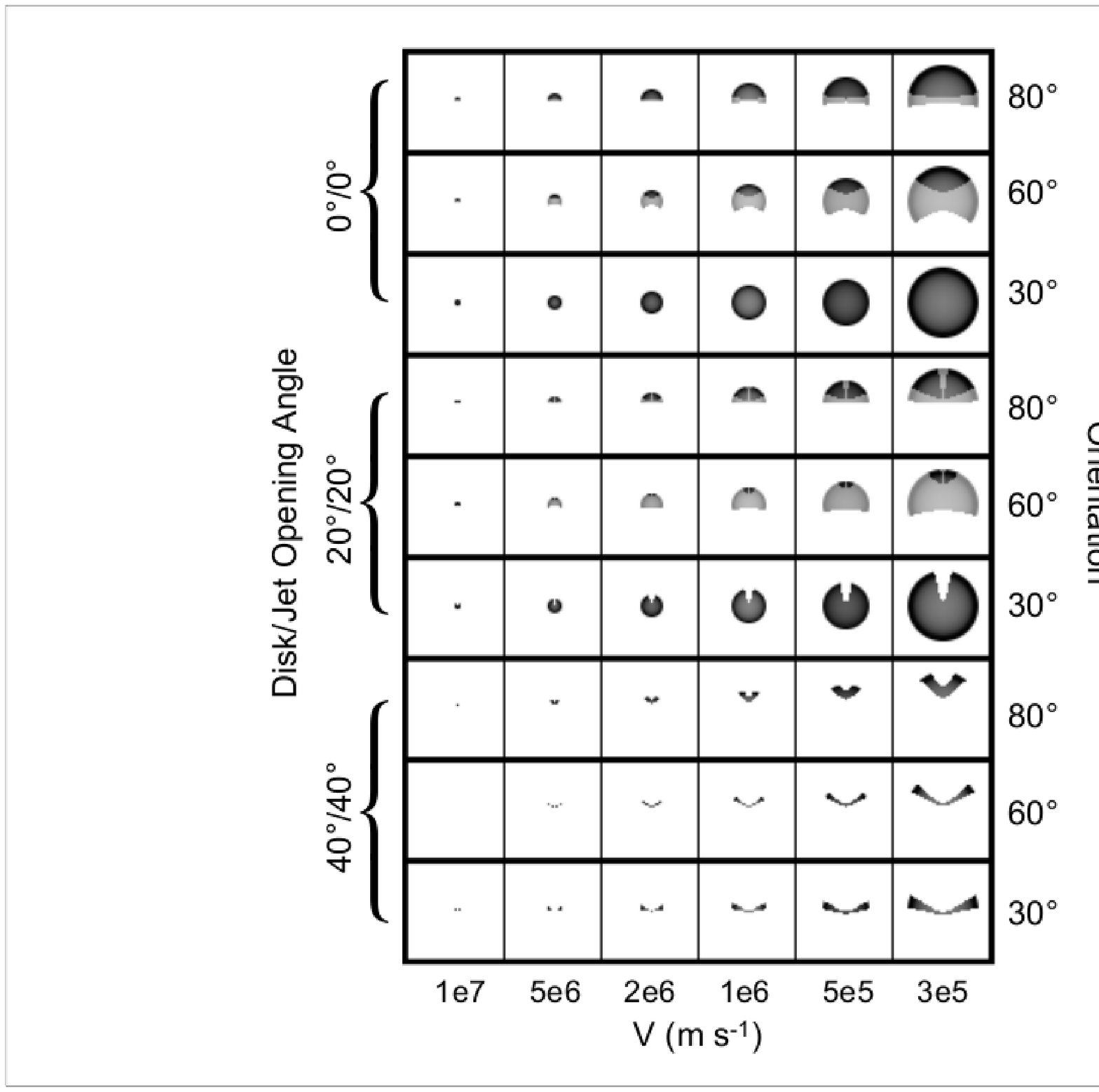}
  \caption{Orbital model surface brightness maps for a selection of
    the explored parameter space. The depicted models have black hole
    mass $m_{BH}=1\times10^{9}$~M$_\odot$ and a range of disk/jet opening
    angles and orientations of the axis to the
    line of sight. For each parameter
    set, a small selection of the calculated velocity slices are
    shown. Receding and approaching gas produces symmetrical surface
    brightness distributions. Each frame size is $3\times10^{17}$~cm.}
  \label{keplsb}
\end{figure}

Figure~\ref{keplsb} shows an example set of surface brightness
distributions for the orbital model. This model produces symmetric
surface brightness distributions for the red- and blue-shifted gas.

\subsection{Parameter Range}

For both the outflow and the orbital model we include obscuration
caused by an equatorial disk and  an axial jet. We explore the
following disk/jet opening angles for these:
$0^{\circ}/0^{\circ}$, $0^{\circ}/10^{\circ}$,
$10^{\circ}/0^{\circ}$ , $10^{\circ}/10^{\circ}$,
$40^{\circ}/40^{\circ}$, $T/0^{\circ}$, $T/10^{\circ}$. {\it ``T''}
indicates a transparent accretion disk; in all other cases the disk is
opaque, meaning that only the near side of the BELR is visible.

For both the kinematic models and all opening angle combinations we
explore the following orientations: 
$0^{\circ}$, $15^{\circ}$, $30^{\circ}$, $45^{\circ}$,
$60^{\circ}$ $70^{\circ}$, $80^{\circ}$, $90^{\circ}$; 

For the outflow model we simulate wind launching radii $r_0$ of
$1\times10^{15}$~cm, $3\times10^{15}$~cm, $1\times10^{16}$~cm, 
$3\times10^{16}$~cm, and $6\times10^{16}$~cm. 
We assume an initial velocity of $v_0=0$ and a terminal velocity for the wind of $V_T=0.1c$ 

For the orbital model we simulate black hole masses $m_{BH}$ of 
$1\times10^{8}$~M$_\odot$, $3\times10^{8}$~M$_\odot$, $1\times10^{9}$~M$_\odot$, 
$1.5\times10^{9}$~M$_\odot$, $2\times10^{9}$~M$_\odot$, $3\times10^{9}$~M$_\odot$, 
$4\times10^{9}$~M$_\odot$, and $6\times10^{9}$~M$_\odot$. We assume a
density gradient of $\alpha=-1$.

For each model and parameter combination we calculate surface brightness
maps for 5\AA-wide wavelength bins around the CIII] central
wavelength between 1809\AA\ and 2009\AA.

\section{Microlensing Simulations}
\label{simulations}

To determine the microlensing signatures predicted by these
BELR models, we perform a Monte Carlo simulation in which we 
convolve the model surface brightness maps with simulated
magnification maps to produce a large set of expected line ratio
spectra. 

The magnification maps are constructed 
using an inverse ray-shooting technique (e.g. \citealt*{k86};
\citealt*{wpk90}). The key parameters for such simulations are the
convergence $\kappa_{tot}$ and the shear $\gamma$ of the lens at the
image positions. The Q2237+0305 lens model is provided in Table
\ref{lensmodel} (Trott \& Wayth, priv. comm., based on models of \citealt{Trott}). We assume a smooth
matter component in the lens of $0\%$, and so the convergence is
provided solely by a clumpy stellar component. This is reasonable as
the lensed images in Q2237+0305 lie very close to the bulge of the
lensing galaxy, where the stellar component is assumed to dominate
(\citealt{kf88}; \citealt{setal88}; \citealt*{swl98}; \citealt*{Trott}). 
A fixed microlens mass of $1\rmn{M}_{\odot}$ was used. The Einstein Radius
changes as the square root of this mass, although clustering of
caustics is relatively unaffected. 

We generated magnification maps for images A and B. These maps have 
$2048 \times 2048$ pixels, with 5 different pixel scales for each.
The scale size corresponds to the scale of the BELR model -- which
increases with $r_0$ and $M_{BH}$ -- and is
chosen to allow sufficient independent positions for the Monte Carlo
simulation (see below). 
Table~\ref{pixelscale} lists the pixel scale used for the different
black hole masses and wind launching radii.

For each BELR model parameter set, we generated surface brightness
maps for 38 wavelength bins spanning 200\AA\ centred on the
emission line central wavelength. At each iteration of the Monte Carlo
simulation we choose two source positions,
one on the image A magnification maps and one on the image B
magnification maps. We then convolve the BELR surface brightness maps
for all 38 wavelength bins with the magnification maps, centred on these
two positions. The resulting pair of magnification spectra yields our
magnificiation ratio spectra for the BELR model, which is equivalent
to a continuum-subtracted line flux ratio.
 
The two magnification map positions are constrained by the observed
continuum flux ratio. We first randomly select a source position
on the B magnification maps. Source positions on the A
magnification map are then randomly selected with the constraint that
the $B/A$ magnification ratio for a point source be within 10\% of the 
observed continuum flux ratio $B/A = 0.39$.

The above assumes that the dominant continuum emission region is
smaller than the spatial resolution element at each pixel scale. 
For the smallest pixel scales 
(map sizes 2.4$\eta_0$ and 7.2$\eta_0$, corresponding to pixel scales of 2.1-6.3$\times10^{14}h_{70}^{-1/2}$~pc),
this gives a continuum size smaller than the upper limit found in Section~\ref{fr}.
However, these sizes are appropriate to the scale of the BELR models considered.
They assume the dominant optical continuum emission arises from a 
region at a factor of $\sim3$ or more smaller than the smallest scale
of  broad-line emission. 

We generated 1215 line ratio spectra per wind model, 4860 line ratio
spectra for the $M_{BH} = 1.5\times10^9 M_{\odot}$, $2\times10^9
M_{\odot}$ and $4\times10^9 M_{\odot}$ orbital models, and 1215 line
ratio spectra for all other orbital models. These line ratio spectra
could then be compared directly with the observed $B/A$ line ratio
spectrum. 

\begin{table}
\caption{Lensing parameters}
\label{lensmodel}
\begin{tabular}{lccl}
\hline
Image & $\kappa_{tot}$ & $\gamma$ & $\mu_{tot}$ \\
\hline
A & 0.413 & 0.382 & 5.034  \\
B & 0.410 & 0.384 & 4.984 \\
\hline
\end{tabular}

\medskip
Lensing parameters for images $A$ and $B$ in Q 2237+0305 (Trott \& Wayth, priv. comm., based on models of \citealt{Trott}).
\end{table}

\begin{table}
\caption{BELR models and magnification maps}
\label{pixelscale}
\begin{tabular}{ccc}
\hline
Map size & orbital models & wind models \\
  & ($M_{bh}$ in $\rmn{M}_{\odot}$) & ($r_0$ in $cm$) \\
\hline
$2.4\eta_0 \times 2.4\eta_0$ & $1\times10^8$ & $1\times10^{15}$\\
$7.2\eta_0 \times 7.2\eta_0$ & $3\times10^8$ & $3\times10^{15}$\\
$24\eta_0 \times 24\eta_0$ & $1\times10^9$, $1.5\times10^9$ & $1\times10^{16}$\\
$72\eta_0 \times 72\eta_0$ & $2\times10^9$, $3\times10^9$, $4\times10^9$ & $3\times10^{16}$\\
$144\eta_0 \times 144\eta_0$ & $6\times10^9$ & $6\times10^{16}$ \\
\hline
\end{tabular}
Magnification map size, and the corresponding orbital and wind
models. $\eta_0$ is the Einstein Radius projected on to the source
plane: $1.79 \times 10^{17} h^{-1/2}_{70} (M/\rmn{M}_{\odot})^{1/2}$~cm 
for Q2237+0305. 
\medskip

\end{table}

\begin{figure*}
  \includegraphics[width=125mm]{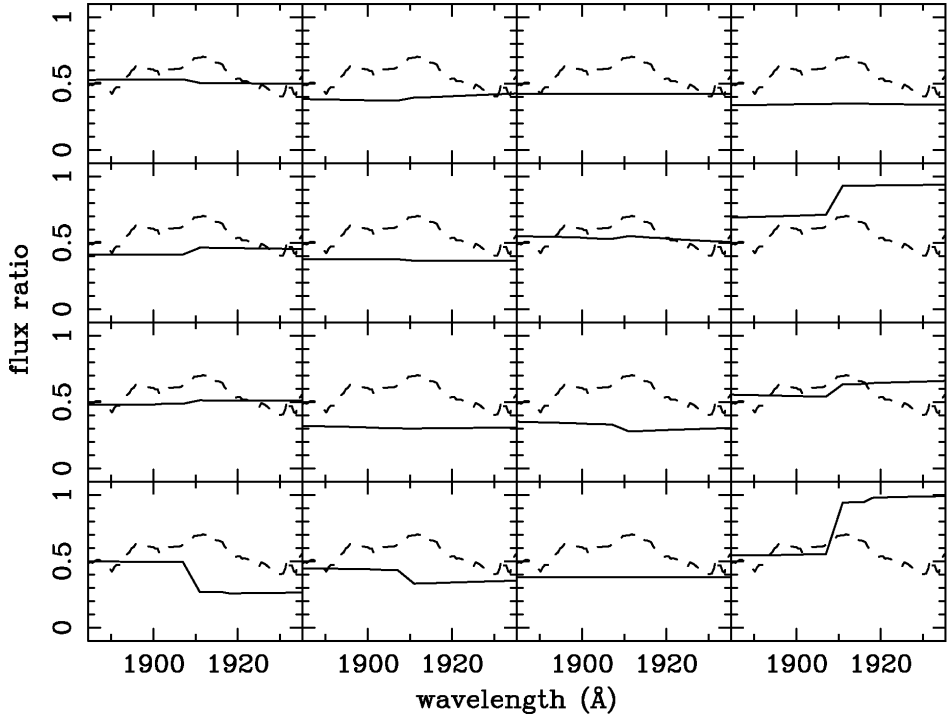}
  \caption{Sample continuum-subtracted line ratios from simulations of
  outflow model (solid lines), with parameters
  $r_0=1\times10^{16}$~cm, orientation 70$\degr$, jet/disk opening
  angle 0$\degr$/0$\degr$. Typical microlensing signatures are
  asymetric, and the observed hump (dashed lines) is 
  rarely reproduced.}
  \label{windmagspec}
\end{figure*}

\begin{figure*}
  \includegraphics[width=125mm]{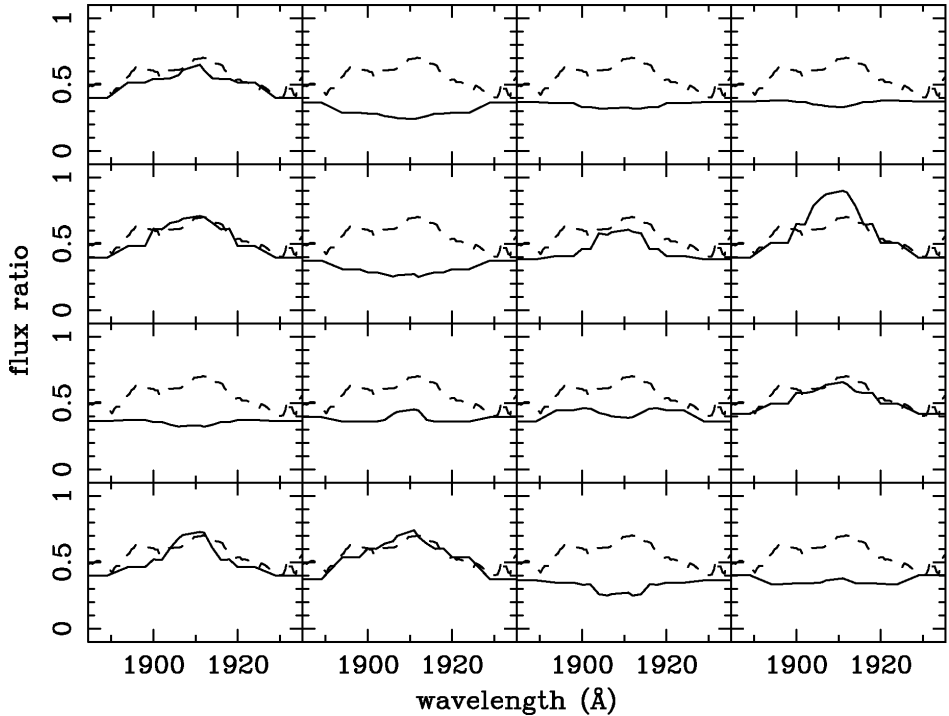}
  \caption{Sample continuum-subtracted line ratios from simulations of
  orbital model (solid lines), with parameters
  $r_0=3\times10^{9}$~M$_\odot$, orientation 70$\degr$, jet/disk
  opening angle 0$\degr$/0$\degr$. The observed microlensing signature
  (dashed lines) is similar to the simulations in many cases.}
  \label{keplmagspec}
\end{figure*}

Figures~\ref{windmagspec} and \ref{keplmagspec} shows a random
selection of simulated magnification ratio spectra for the outflow model
and the orbital model respectively, for a single parameter
combination. For outflow model the parameters are:
$r_0=1\times10^{16}$~cm, orientation 70$\degr$, disk/jet opening angle
0$\degr$/0$\degr$. For the orbital model the parameters are: 
$M_{BH}=3\times10^{9}$~M$_\odot$, orientation 0$\degr$, disk/jet opening angle
0$\degr$/0$\degr$.

\section{Measurement of BELR Kinematics}

\begin{figure*}
  \includegraphics[width=100mm]{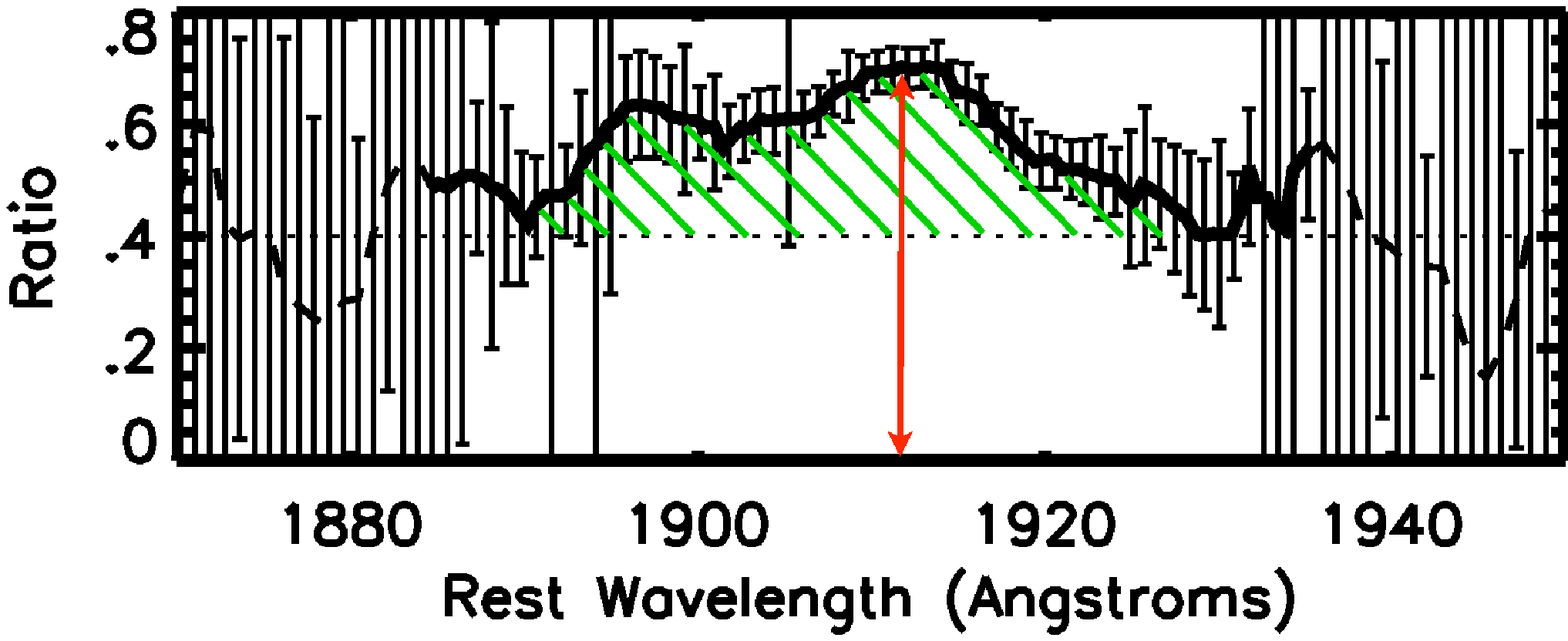}
  \caption{Parameterization of the shape of the observed CIII]
    microlensing signature: we define $W$, an equivalent width analog for
    the ratio of continuum-subtracted line ratio for images B/A:
    the integrated ratio minus the continuum ratio ({\it green area}) 
    divided by $L$, the central-wavelength amplitude of the ratio
    ({\it red arrow}).
}
  \label{eqw_method}
\end{figure*}

As illustrated in Figures~\ref{windmagspec} and ~\ref{keplmagspec}, the
orbital model tends to reproduce the microlensing signature observed
in the data more easily than the wind model. 
To quantify this comparison, we
adopt a simple parameterization for the observed microlensing
signature, comparing the amplitude and width of the observed hump. 
Taking the CIII] continuum-subtracted line ratios between
images A and B, we compare the amplitude at the rest 
wavelength of the emission line (designated $L$) to an equivalent
width analog: the integral across the line's width minus the continuum
ratio, divided by $L$ (designated $W$). This is illustrated in figure~\ref{eqw_method}.
This parameterization is useful because it is sensitive to the
central amplitude, the integrated strength, and the symmetry of the
feature, which together are effective in identifying shapes similar to
the observed microlensing signature.

\begin{figure*}
  \includegraphics[width=170mm]{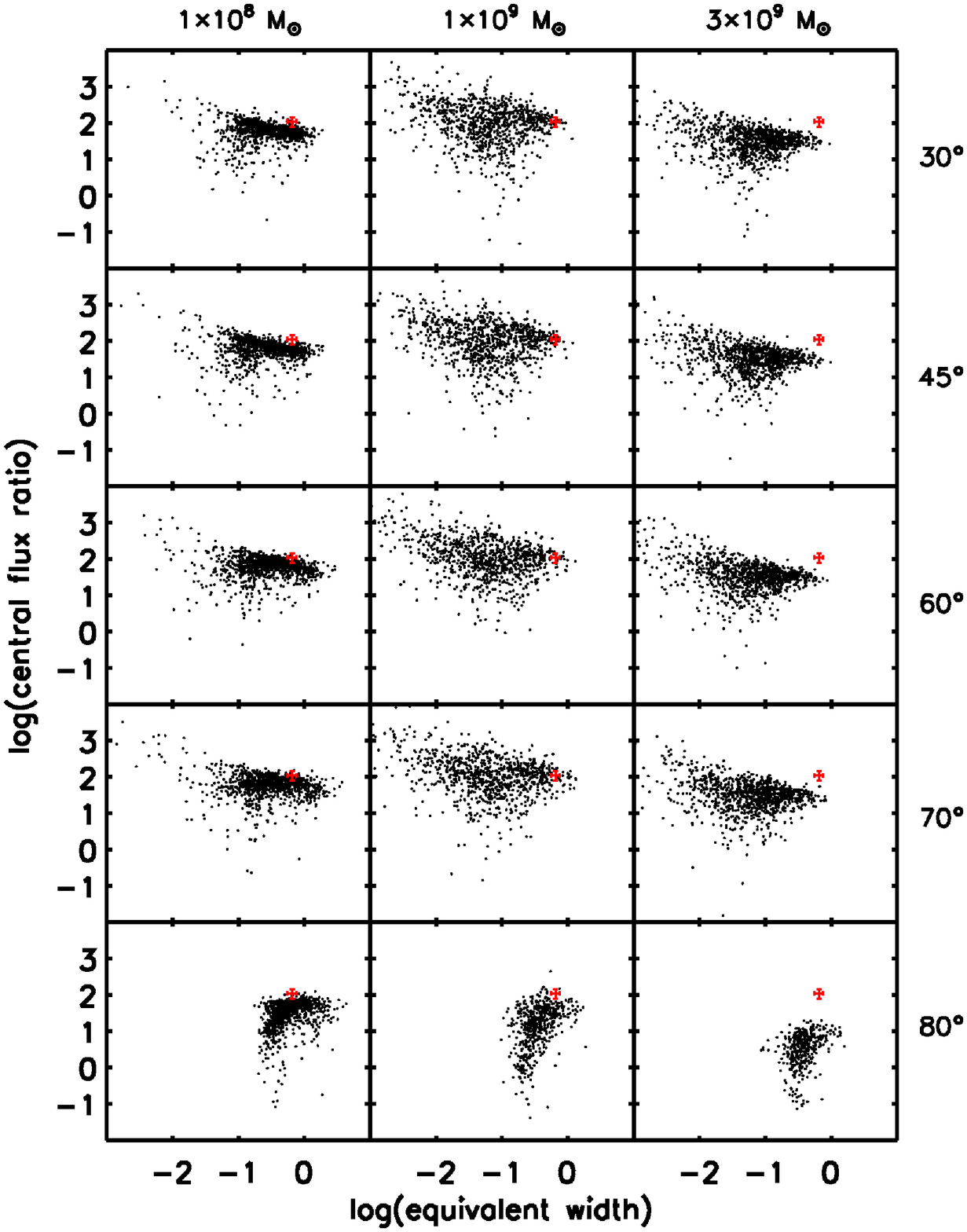}
  \caption{``Equivalent width'' of the continuum-subtracted line ratio
    versus the central flux ratio for simulations ({\it black points})
  and data ({\it red error bars}). A selection of the explored
  parameter space is shown: 
  M$_{BH}$ of $1\times10^{8}$~M$_\odot$ to $3\times10^{9}$~M$_\odot$,
  orientations of 30$\degr$ to 80$\degr$, and jet/disk opening angle of 
  20$\degr$/20$\degr$. A black hole mass of $1\times10^{9}$~M$_\odot$ is
 favoured, while a black hole mass of $3\times10^{9}$~M$_\odot$
 appears unlikely.
}
  \label{lrrlocikepl}
\end{figure*}

\begin{figure*}
  \includegraphics[width=170mm]{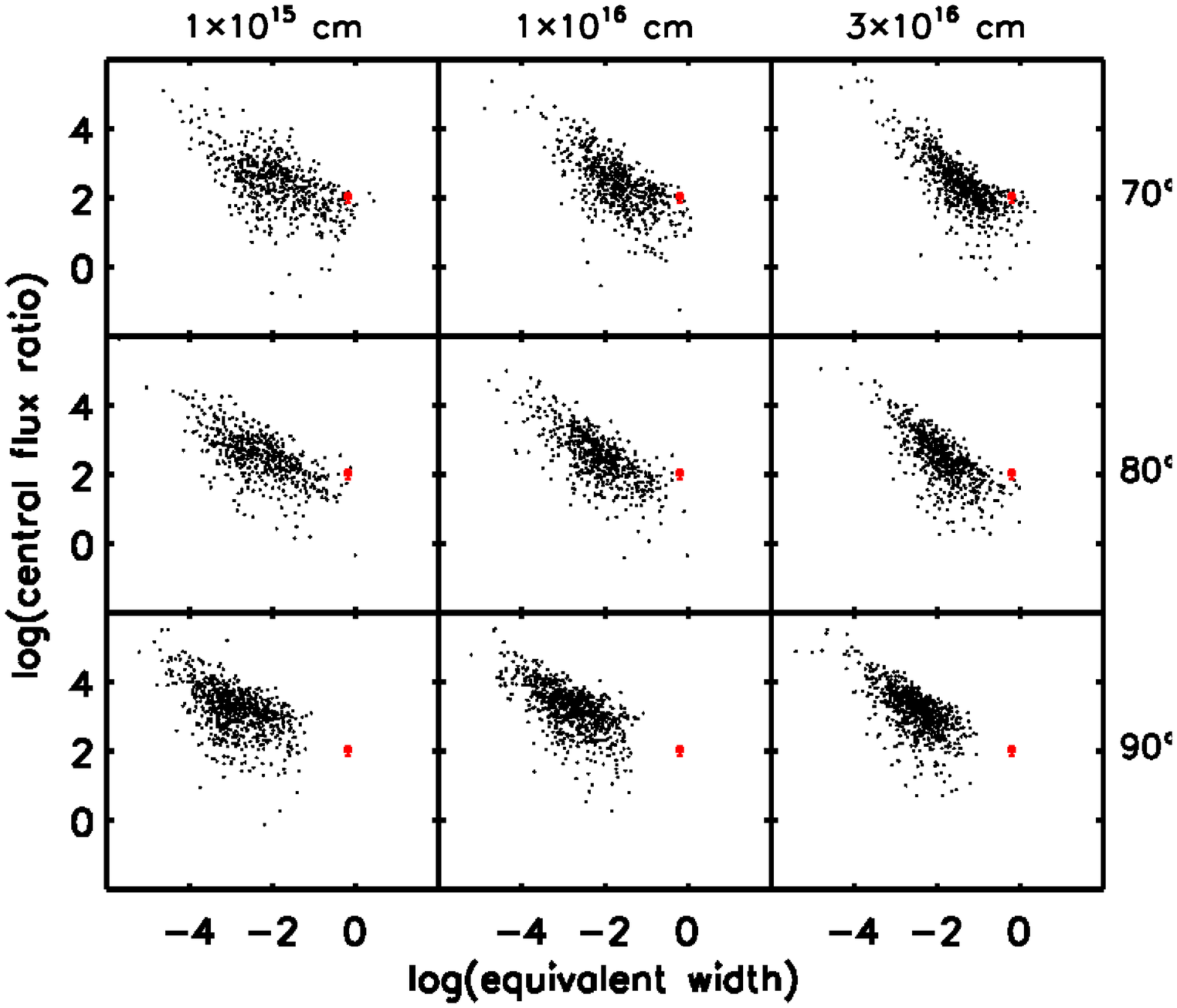}
  \caption{``Equivalent width'' of the continuum-subtracted line ratio
    versus the central flux ratio for simulations ({\it black points})
  and data ({\it red error bars}). A selection of the explored
  parameter space is shown: 
  r$_{0}$ of $1\times10^{15}$~cm,  $1\times10^{16}$~cm,  and $3\times10^{16}$~cm,
  orientations of 70$\degr$, 80$\degr$, and 90$\degr$, and jet/disk opening angle of 
  20$\degr$/20$\degr$. For smaller orientations the data point falls
  much further from the simulations.}
  \label{lrrlociwind}
\end{figure*}

Figures~\ref{lrrlocikepl} and \ref{lrrlociwind} show $L$ versus $W$ for a selection of
the parameter space explored in the simulations. The error bars show
the location of the data in this parameterization and the
$2\sigma$ uncertainty. 

Using this parameterization, we can calculate the p-value for the observed
microlensing signature with respect to each model. The probability, $P$, of
each model reproducing the observed microlensing signature within 
$2\sigma$ of $L$ and $W$ is the ratio of the number of simulated points
within these error bars to the total number of simulations. 
The p-value is then the integrated probability distribution below
$P$, which is equal to 
the number of simulation points with a surrounding point density within a
$2\sigma$ box equal to or lower than the point density of the
observed signature, divided by the total number of simulation points. 
Given these p-values, we can reject models and parameter
combinations with a quantifyable significance level.

Figure~\ref{eqwprobkepl} shows the p-values for the orbital model as
a function of orientation, for a range of disk and jet opening angles
and black hole masses/launching radii. Fig.~\ref{eqwprobkepl} 
(right) shows the average probabilities over all orientations as a
function of black hole mass/launching radius.  

\begin{figure}
\hspace{-1cm}
  \includegraphics*[angle=-90,width=100mm]{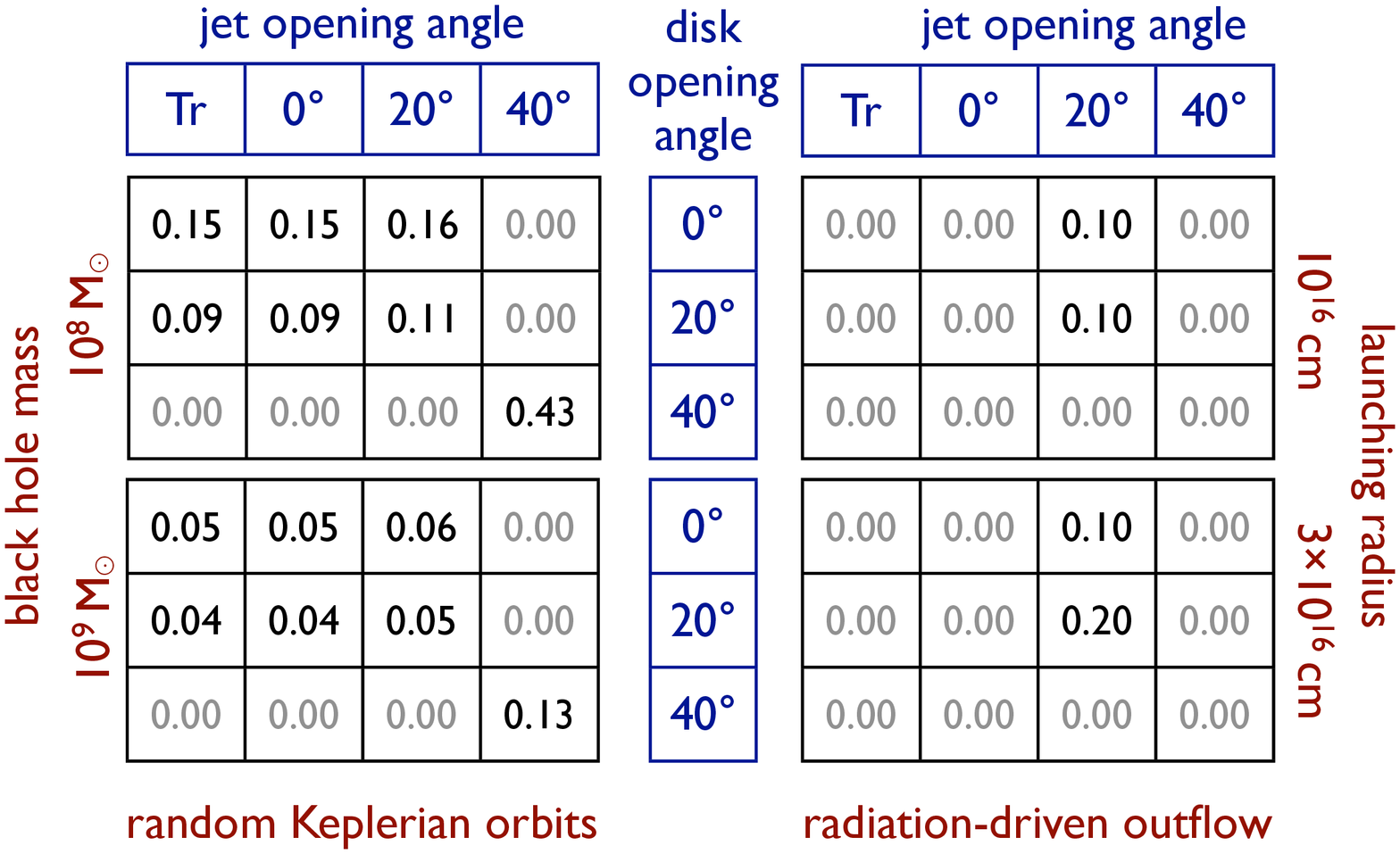}
  \caption{Probability of the observed microlensing signature as a
    function of model parameters. These values are derived by
    integrating the density functions of the simulated central ratio
    vs. equivalent width distributions 
    (Figs.~\ref{lrrlociwind} and ~\ref{lrrlocikepl}) 
    up to the density found at the location of the observed signature.
    {\it Left}: orbital model, with probabilities averaged over all
    orientations. {\it Right}: outflow model, with probabilities
    averaged over 70$\degr$ and 80$\degr$ orientations
    only. For all other orientations the outflow model produced
    no results close to the observed signature for any launching radius.
}
  \label{probtable}
\end{figure}

Figure~\ref{probtable} shows the orientation-averaged p-values for
both the orbital and the 
wind models, for all disk and jet opening angles and a selection of
black hole masses and wind launching radii.

Assuming the orbital model, the p-value for the observed
microlensing signature is inversely proportional to black hole mass. 
We can place a firm upper limit on black hole mass, assuming this
orbital model. Models 
with $M_{BH} \ge 1.5\times10^9 M_{\odot}$ are rejected with $>$99\%
confidence. Larger black hole masses are ruled out for all
orientations and disk/jet opening angles. 

The most probable disk/jet configuration for all orientations and
black hole masses is one with a large opening angle for both disk and
jet of $40\degr$. Interestingly, cases where only one of the disk or
the jet have a large opening angle are ruled out with $>$99\%
probability.

For the pure outflow model, very few parameter sets are capable of reproducing
the observed microlensing signature. All are ruled out with high
($>99\%$) confidence with the exception of models with very small
starting radii for the wind ($10^{15}$~cm; probably unphysical as it
is smaller than the lower estimates of the accretion disk size,
e.g. \citealt{Eigenbrod08b}) and
a tightly constrained orientation range ($\sim70\degr$), which are
excluded at only the 90\% level. This
tells us that it is unlikely that a pure accelerating outflow produced the
observed microlensing signature. 

Changing the mean microlensing mass marginally affects the p-values,
but the general results are the same. If mean masses are smaller,
differential microlensing becomes less likely but increases in
strength, and vice versa for larger mean masses. It becomes no easier
to produce the observed microlensing signature with a pure outflow
model.

\section{Discussion}

Reverberation mapping studies have now provided evidence for 
inflowing, outflowing, Keplerian and virialised motions in quasar BELRs
\citep{Denney09, Bentz09}.  Using a wide range of observational
evidence, \citet{Gaskell09} has argued in favour of infall velocities
which are smaller than the Keplerian velocities.  Flow-dominated BELR
models, for example those outlined by \citet{br91} and \citet{elvis} are
especially attractive because it is known that high velocity nuclear
flows are very common, if not ubiquitous, in quasars.  Indeed, the fraction of
quasars in SDSS showing some CIV broad absorption is ~25\%
\citep{Trump06}.  The line-driven photoionised wind
modelled by \citet{Murray95} produced line profiles in good agreement
with observation.  In addition, there are now strong physical
arguments for circular motions (gravity), and also the first
observational evidence  from polarisation of the H$\beta$ line 
\citep{Young07}.  Thus there is still no consensus on even the
simple question of the direction of the gas flow in the broad line
region.

Our analysis shows that a pure Keplerian model for the BELR is
able to reproduce the observed microlensing signature. In
contrast, our pure accelerating outflow model cannot reproduce this
signature for any physical launching radius. 
This is expected considering the qualitative nature of this
signature. The low velocity gas exhibits the strongest differential
microlensing relative to the continuum (which is expected to arise
from the smallest size scale), while the high velocity gas exhibits
flux ratios closer to that of the continuum. 
This distribution of
velocities is consistent with a Keplerian field in which the highest
velocities are expected closest to the central gravitating body, but
inconsistent with a purely accelerating outflow in which velocity
increases with distance.

The observed differential microlensing signature for the BELR is also
consistent with that observed blueness of the continuum for image A: 
if smaller size scales are preferentiallly magnified, then we expect the continuum to
appear more blue, assuming that this magnification gradient extends to
scales smaller the outer regions of the accretion disk. 


\subsection{Implications for Outflow Models}

The observed differential microlensing signature rules out BELR models
in which the gas dynamics at all radii are dominated by an accelerating outflow. 
This includes winds that are exclusively driven by line pressure.
Other drivers, such as thermal expansion of the 
confining gas \citep{Weymann82}, centrifugal force
\citep{BlandfordPayne82, Emmering92} or radiation pressure via dust 
\citep{Voitetal93} may produce outflows in which
high velocity gas dominates at smaller radii.

Even if the driving force behind the outflow only allows for an
accelerating wind, such models are still viable so long as the gas
remains within the gravitationally dominated region for a period
comparable to or greater than its orbital period.

\subsection{A Pure Keplerian Field?}

Until recently, reverberation mapping studies have found that BELR
velocity distributions are consistent with Keplerian motion \citep{PetersonWandel00}. Our
results are consistent with this, and extend the finding of
reverberation mapping---itself applicable to 
low-luminosity AGN---into the high-luminosity regime.
Gravitationally-dominated kinematics are also supported by the fact that 
the AGN black hole masses calculated based on the assumption of
a virial BELR follow the $M_{BH}$-$\sigma_*$ relation of normal
elliptical galaxies \citep{Ferrareseetal01}.

Recent reverberation mapping results with greatly improved velocity
resolution have now found evidence for both infall- and
outflow-dominated BELR kinematics \citep{Denney09, Bentz09}, 
indicating that a variety of dynamical processes are important.
Simple randomly oriented orbits are unlikely.
It is known that the inner region of quasar BELRs can be as
small as $10^{16}$~cm \citep{Kaspi00}, which overlaps the outer
regions of accretion disk. This greatly restricts the possible orbits in
the innermost regions of the BELR.

Given the likely presence of multiple kinematic regimes within the
BELR, we interpret our results cautiously. Rather than implying a
spherical Keplerian field, we instead take these results to indicate
that the inner regions of the BELR are strongly gravitationally dominated, 
which may include orbital motion and/or infall.

\subsection{A Constrained Vortex?}

In our simple models,
the configuration that most easily reproduces the observed
microlensing signature is one in which both accretion disk and jet have
large ($40^\circ$) opening angles; configurations in which only the
disk {\em or} jet have a large opening angle are ruled out with high
confidence. Interestingly, 
this configuration of large opening angles is similar to the
hollowed-cone geometry of many wind models. Thus, our
modeling suggests that such a geometry is viable, but {\em only} if
the gas motion within this outflow cone is gravitationally dominated
at small radii. This is consistent with the vortex model of
\citet{elvis}, in which line-driven gas outflowing in a hollowed cone initially
orbits at accretion disk velocities, or with the
magnetically-constrained outflow similar to that proposed by
\cite{Emmering92}.

This interpretation is also consistent with the biconical model
proposed by \citet{Abajas07} to explain the recurrent microlensing
enhancement of the broad line blue wings in SDSS J1004+4112.

\subsection{Black Hole Mass Limit}

Our black hole mass limit of $M_{BH} < 1.5\times10^9 M_{\odot}$ is
of similar order to the virial estimate based on the width of the CIV
line ($0.9\times10^9  M_{\odot}$; \citealt{Morgan10}). 
If we calculate a virial black hole mass from the upper limit on the
CIII] BELR from Section~\ref{fr} and the line's velocity FWHM, we
instead obtain $M_{BH} < 2\times10^8 M_{\odot}$. 

Modeling the differential microlensing signature with a BELR comprised of a simple
Keplerian field seems to be as useful for obtaining black hole mass limits
based on the virial assumption as the standard method. It has the added
advantage of including both BELR velocity and size in the model,
instead of using quasar luminosity as a proxy for BELR size 
(calibrated using low-luminosity quasars). However stronger
limits are obtainable by placing a more direct limit on BELR size from
microlensing of this region. 

\subsection{Comparison to Other Works}

\citet{ecsma} also detect differential microlensing within the velocity
structure of the BELR in their long-running VLT monitoring campaign.
The nature of the signatures are broadly consistent with those we
observe; in high-magnification phases, the wings of the CIII] line were
more strongly amplified than the core.

\begin{figure}
  \includegraphics*[width=85mm]{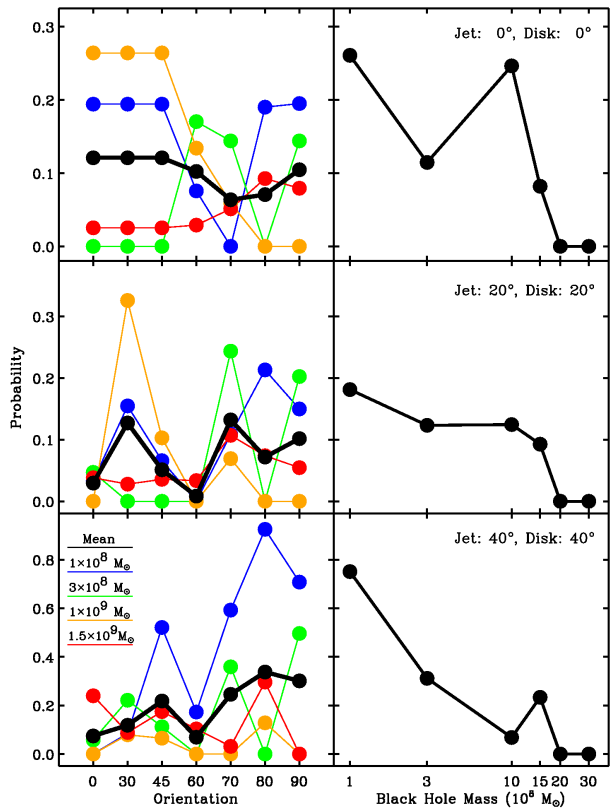}
  \caption{{\it Left:} p-value as a function of orientation for a
    range of black hole masses 
({\it blue:} 10$^8$~M$_\odot$, {\it green:} 3$\times$10$^8$~M$_\odot$, {\it yellow:} 10$^9$~M$_\odot$, {\it red:} 1.5$\times$10$^9$~M$_\odot$), 
   and the average 
    p-value over black hole masses ({\it black lines}). {\it Right:}
    average p-value over all orientations as a function of black hole mass.
    Vertical panels show results for different jet/disk opening angles 
({\it upper:} 0$\degr/$0$\degr$, 
{\it middle:} 20$\degr/$20$\degr$, 
{\it lower:} 40$\degr/$40$\degr$).
}
  \label{eqwprobkepl}
\end{figure}

\citet{PK} are able to constrain the orientation of the accretion disk
of Q2237+0305 by comparing the expected light curve produced by a
model accretion disk with the continuum light curve observed over 11
years of OGLE monitoring. Their finding of $cos~i > 0.66$, which is to
say a face-on accretion disk within $\sim 50\deg$ of the line of
sight, is consistent with our results, which don't strongly constrain
orientation except for a few specific parameter combinations; in
particular, Keplerian models with the largest black hole masses (see Fig.~\ref{eqwprobkepl}).

\section{Conclusion}

We have presented observations of the gravitationally lensed quasar
Q2237+0305 taken with the Gemini North GMOS IFU. Ratios between
continuum subtracted image B and image A spectra reveal differential
microlensing across the velocity structure of the CIII] broad emission
line. The high velocity wings of this line tend towards the flux ratio
of the continuum, and the lower velocity core, while still
microlensed, is closer to the expected flux ratio in the absence of
microlensing. This implies that the high velocity component is emitted
from a region with a size comparable to that of the continuum emission
region, whereas the low velocity component is emitted from a larger
region. 

We conducted microlensing simulations using two simple models of the
broad emission line region: an outflow model and an orbital model. The
outflow model assumes a clumpy wind accellerated by radiation pressure.The
orbital model assumes circular Keplerian orbits with random
orientations. 
For both models we tested a wide range of parameters. These models
were used to construct an 
ensemble of simulated B/A flux ratios as a function of velocity, for
comparison with the observed flux ratio spectrum. 

A purely radial outflow was unable to reproduce the observed
differential microlensing signature for any plausible launching radius
of the wind. Conversely, the orbital model was able to reproduce the
observed signature for all simulated black hole masses $M_{bh}<
2\times10^9M_{\odot}$. Though our orbital model is simplistic, we
interpret this result as further evidence that the inner regions of
the BELR are gravitationally dominated. This is consistent with an
outflow model where the BELR gas is lifted off the quasar accretion
disk, and thus retains a high Keplerian velocity. 

The BELR models presented here are not intended to accurately describe the physical situation in the quasar. They describe only the generic behaviour of a radially outflowing wind, or a collection of orbiting clouds. More sophisticated models, perhaps making use of radiative transfer codes such as CLOUDY, may provide additional constraints on the quasar central engine. 

We have obtained Gemini IFU observations of nine other gravitationally lensed quasars, both double and quadruply imaged. These data will allow us to probe a range of quasar orientations, emission lines, black hole masses, and therefore BELR emission region scales. The analyses of these data are forthcoming.

\section{Acknowledgements}
NFB acknowledges the support of an Australian Postgraduate Award.


\bsp

\label{lastpage}

\end{document}